\documentclass[12pt]{iopart}
\usepackage{graphicx}% Include figure files
\usepackage{dcolumn}% Align table columns on decimal point
\usepackage{color}
\usepackage{bm}% bold math
\usepackage{amssymb}
\usepackage{amsfonts}
\usepackage{epsf}
\usepackage{epsf}

\newcommand {\beq}{\begin{equation}}
\newcommand {\eeq}{\end{equation}}
\newcommand {\bea}{\begin{eqnarray}}
\newcommand {\eea}{\end{eqnarray}}

\begin{document}

\title
{Inhomogeneous Coupling in Two-Channel Asymmetric Simple Exclusion Processes}
\author{K.Tsekouras\dag  and  A.B.Kolomeisky\ddag}
\address{\dag Department of Physics, Rice University, Houston, TX 77005-1892}
\address{\ddag Department of Chemistry, Rice University, Houston, TX 77005-1892}

\begin{abstract}\\
Asymmetric exclusion processes for particles moving on parallel channels with inhomogeneous coupling are investigated theoretically. Particles  interact with hard-core exclusion and move in the same direction on both lattices, while transitions between the channels is allowed  at one specific location in the bulk of the system. An approximate theoretical approach that describes the dynamics in the vertical link and horizontal lattice segments exactly but neglects the correlation between the horizontal and vertical transport is developed. It allows us to calculate  stationary phase diagrams, particle currents and densities for symmetric and asymmetric transitions between the channels. It is shown that in the case of the symmetric coupling there are three stationary phases, similarly to the case of single-channel totally asymmetric exclusion processes with local inhomogeneity. However, the asymmetric coupling between the lattices lead to a very complex phase diagram with ten stationary-state regimes. Extensive Monte Carlo computer simulations generally support theoretical predictions, although simulated stationary-state properties slightly deviate from calculated in the mean-field approximation, suggesting the importance of correlations in the system.  Dynamic properties and phase diagrams are discussed by analyzing constraints on the particle currents across the channels. 
\end{abstract}

\pacs{05.70.Ln,05.60.Cd,02.50Ey,02.70Uu}

\ead{tolya@rice.edu}

\submitto{\JPA}

\maketitle

\section{Introduction}

In the absence of fundamental framework for  non-equilibrium systems, one-dimensional models known as asymmetric simple
exclusion processes (ASEP), where particles interact via an exclusion potential and hop along discrete lattices, have become
important tools and references for  understanding complex  phenomena in chemistry, physics and biology
\cite{derrida98,schutz}. ASEPs now are playing a role similar to the one the Ising model has in the studies of equilibrium systems
\cite{schutz03}. Asymmetric exclusion processes have been applied successfully to analyze the kinetics of biopolymerization \cite{macdonald68}, protein synthesis \cite{shaw03,shaw04,chou04}, molecular transport through nanopores and channels \cite{chou99}, the motion of
motor proteins along cytoskeleton filaments \cite{klumpp05,parmeggiani03,nishinari05}, growth of fungal filaments \cite{evans07}, and different aspects of car traffic processes \cite{helbing01,chowdhury00}. 

Many non-equilibrium phenomena can be reasonable well described by a single-channel exclusion models. However, the necessity to analyze more realistic complex phenomena, that involve the transport along the parallel coupled lattices and existence of internal states, e.g. for motor proteins, vehicular traffic and hopping of quantum dots, stimulated the development of multi-channel ASEPs \cite{popkov01,pronina04,pronina05,mitsudo05,harris05,pronina06,reichenbach06}. In these models particle can move along their lattice and they also can transfer between the channels. leading to complex dynamic behavior. Theoretical analysis of multi-channel exclusion processes indicates that a coupling between channels has a strong effect on particle dynamics and stationary-state properties, and it also leads to several unusual phenomena, such as localized domain walls \cite{pronina06,reichenbach06,pronina07}.

Dynamics of many non-equilibrium systems is strongly modified by the presence of inhomogeneities and defects. For example, it is known that in the protein synthesis the elongation rate of ribosomes is not uniform due to slow codon sites on RNA molecules \cite{chou04,solomovici97}. The effect of local inhomogeneities  have been already investigated for different ASEPs on single-chain lattices \cite{shaw04,kolomeisky98,mirin03,ha03}. However, for multi-channel ASEPs mostly homogeneous systems have been considered. The aim of the present paper is to investigate two-channel ASEPs with inhomogeneous coupling between the lattices. Specifically, we analyze the simplest case when two channels are coupled at the single lattice site far away from the boundaries, and two different cases, symmetric and asymmetric couplings, are considered.   The problem is motivated by the cellular transport of  motor proteins that move along parallel protofilaments along microtubules \cite{howard}. Occasionally, motor proteins  may jump between the chains, especially in the presence of structural defects in the original protofilament. In our study we use a simple approximate model for  theoretical calculations of stationary-state properties and phase diagrams,  and we test our predictions by utilizing extensive Monte-Carlo simulations. 

The paper is organized as follows. In section 2 we give a detailed description of the model, briefly present known results for single-chain ASEPs, and develop an approximate theory for asymmetric exclusion processes involving two channels connected by a single cluster far away from the boundaries  where particles can transfer with symmetric or asymmetric rates between the lattices.  In section 3 we present and discuss Monte-Carlo computer simulation results and compare them with theoretical predictions.   Finally, in section 4 we summarize and conclude.

\section{Theoretical description of  two-channel ASEPs with inhomogeneous coupling}

\subsection{Model}

We consider identical particles that move on two parallel one-dimensional lattices as illustrated in figure
1. Two channels are identical and they have $L+1$ sites. The particles  interact via hard-core exclusion potential:  each site is either empty or occupied by a single particle.  The dynamics of the system is random-sequential, i.e., at each time step we randomly choose a site on one lattice to follow its dynamics.  Particles can enter the system from the left with the rate $0 < \alpha \le 1$ if any of the first sites in either lattice are empty. Particles that reach the final site in either lattice can leave the system  with the rate $0 < \beta \le 1$. In the bulk of the system particles move  from the site $i$ to the site $i+1$ on the same lattice with the rate 1 if the next site is available. However, dynamic rules are different at the  special site $L/2+1$: see figure 1. A particle at this site on the first lattice can move vertically to the site $L/2+1$ on the second lattice with the rate $w_1$ if the corresponding site  is empty, or it can hop to the right empty  site with the rate $(1-w_{1})$. However, if the site $L/2+1$ in the second channel is occupied, the particle jumps in the horizontal direction to the right with the rate 1 if the forward site is available. The total transition rate out of this defect site is equal to one. Similarly, the particle on the site $L/2+1$ on the second lattice can hop vertically with the rate $w_{2}$ if the upper site  is free, or it can move  horizontally with the rate $(1-w_{2})$ if the forward right site is not occupied. Otherwise, it will move to the right neighboring site with the rate  1, assuming again that the forward site is available. When the transition rates between the channels are equal ($w_{1}=w_{2}$) the coupling is symmetric, while for $w_{1} \ne w_{2}$ the coupling between the lattices is asymmetric. Both cases of two-channel ASEPs with inhomogeneous coupling are analyzed separately.

 \begin{figure}[h]
\centering \includegraphics[scale=0.6,clip=true]{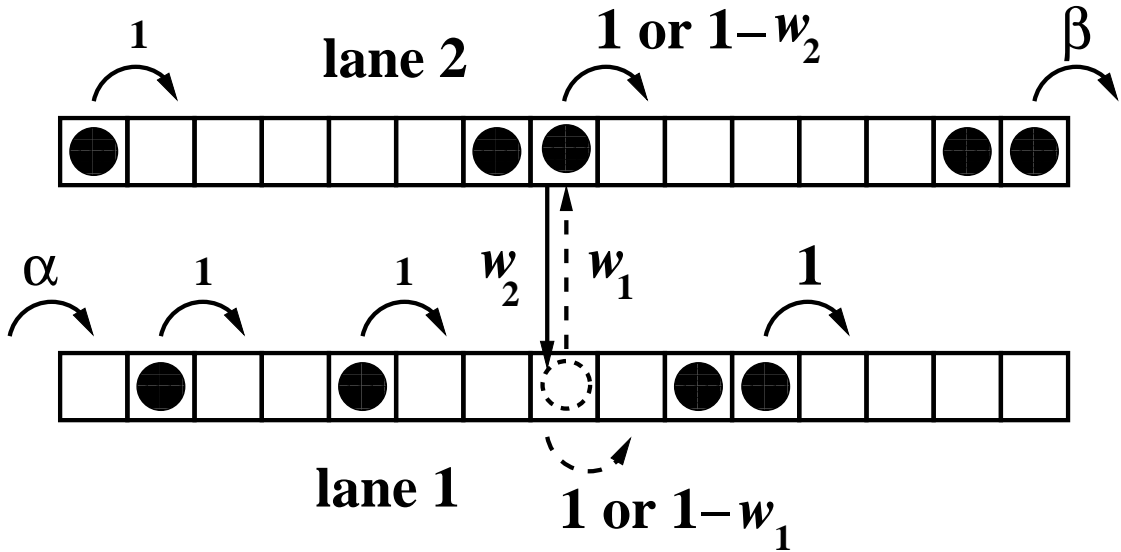} \caption{Schematic view of two-channel ASEP with a coupling at the single  site. Particles move only along the lattices, except on the link site where they can hop to the other lane if the corresponding site is empty. The inter-channel transitions rates are $w_{1}$ and $w_{2}$.  Allowed transitions are shown by arrows. Entrance rates at both lattices are equal to $\alpha$ and exit rates are equal to $\beta$.}
\end{figure}

If the inter-channel hopping rates are equal to zero ($w_{1}=w_{2}=0$) then we have two independent single-lane ASEPs for which exact solutions for all dynamic properties are already known \cite{derrida98}. There are three stationary-state phases.  For large entrance and exit rates, $\alpha>1/2$ and $\beta>1/2$, the single-lane ASEP is found to be in a maximal current (MC) phase where the stationary current and bulk density are given by
\begin{equation}\label{mc1}
J_{MC}=\frac{1}{4}, \quad \rho_{bulk,MC}=\frac{1}{2}.
\end{equation}
For $\alpha<1/2$ and $\alpha<\beta$ the dynamics is determined by the entrance process and we have a low-density (LD) phase where
 the particle current and the bulk density are 
\begin{equation}\label{ld1}
J_{LD}=\alpha(1-\alpha), \quad \rho_{bulk,LD}=\alpha.
\end{equation}
Finally, for $\beta<1/2$ and $\beta<\alpha$  the exiting is a rate-limiting process, and the single-chain ASEP is in a  high-density (HD) phase where
\begin{equation}\label{hd1}
J_{HD}=\beta(1-\beta), \quad \rho_{bulk,HD}=1-\beta. 
\end{equation}

Several approximate theoretical approaches have been developed for investigation of multi-channel ASEPs \cite{popkov01,pronina04,pronina05,mitsudo05,harris05,pronina06,reichenbach06,pronina07}. However, probably the closest description of non-equilibrium dynamics, as judged by comparison with Monte Carlo computer simulations, is obtained in the mean-field model that have  exact treatment of the  vertical transitions between the channels \cite{pronina04,pronina06}. We adopt this strategy to analyze the inhomogeneous coupling in two-channel ASEPs. The dynamics at the coupling cluster can be specified by introducing functions $P_{ij}$ ($i,j=0,1$) that define the probability that the vertical cluster is  empty ($P_{00}$), partially occupied [$P_{10}$ or $P_{01}$ if the occupied site is on the lane 1 or 2, respectively], or fully occupied  ($P_{11}$).  These probabilities are related via a normalization condition  
\begin{equation}
P_{00}+P_{10}+P_{01}+P_{11}=1.
\end{equation}
In the presented analysis we calculate only bulk densities in all phases, however, our analysis can be easily applied to obtain full density profiles utilizing known results \cite{derrida98}.

\subsection{Symmetric coupling}

Let us consider  the case of the symmetric coupling when the vertical transitions rates are equal. As can be seen from figure 1, the vertical coupling cluster divides the two-channel inhomogeneous system into four homogeneous horizontal segments. Each of these segments can be viewed as a single-lane ASEP that are connected with the vertical junction cluster, as shown in figure 2. We label left segments in the first and second channels as 1L and 2L, respectively. The particles enter these segments with the rate $\alpha$ and leave them with an effective rate $\beta_{eff}$. Similarly, 1R and 2R describe the corresponding right segments with entrance and exit rates given by $\alpha_{eff}$ and $\beta$: see figure 2.  Because of the symmetry, the probability to have the partially filled vertical cluster with the particle on lane 1 or 2 is the same, $P_{10}=P_{01}$, which modifies the normalization condition,
\begin{equation}\label{norm_sym}
P_{00}+2P_{10}+P_{11}=1.
\end{equation}
Thus dynamics in four segments and in the vertical cluster is treated exactly, but correlations between different parts are neglected.

\begin{figure}[h]
\centering \includegraphics[scale=0.6,clip=true]{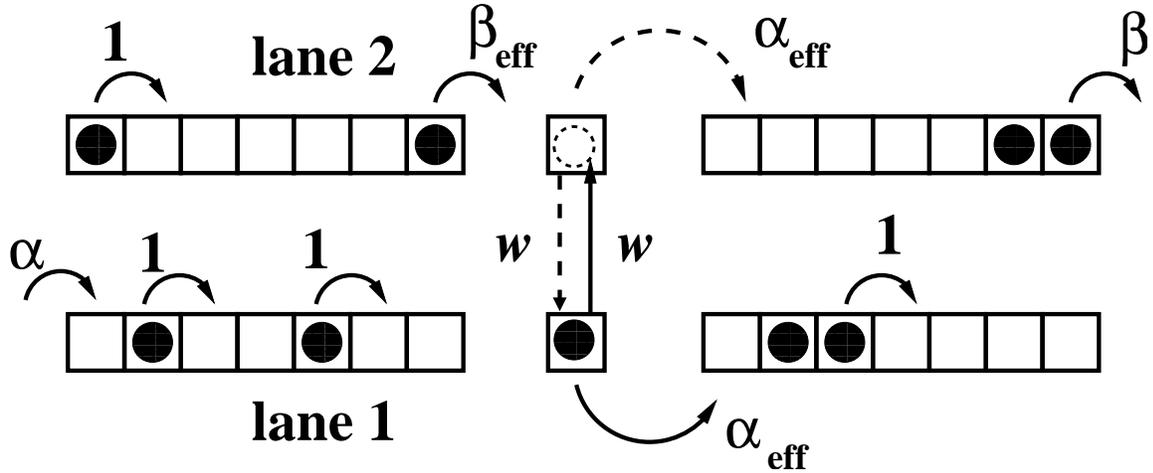} 
\caption{Effective scheme for two-channel ASEPs with inhomogeneous symmetric coupling. The system consists of four segments coupled by a vertical junction cluster. Each segment is viewed as a single-chain ASEP. Allowed transitions are shown by arrows.}
\end{figure}

It is instructive to analyze first the simplest case of strong symmetric coupling when $w_1=w_2=1$. It can be shown that it is not possible to have a completely empty vertical cluster in the large-time limit, i.e., $P_{00}=0$. The vertical configuration ${00}$ can only be obtained from the partially filled cluster by particles leaving to segments 2L and and 2R. However, the first particle to reach the junction cluster will hop vertically from one defect site to the other constantly until another particle arrives to completely fill the cluster. There is no possibility for creating empty vertical cluster. Then from the normalization condition (\ref{norm_sym}) one obtains,
\begin{equation}
P_{10}=\frac{1-P_{11}}{2}.
\end{equation}

Because  from the vertical cluster particles can leave to the right part of the system only from the completely filled configuration, the effective rate to enter the segments 1R and 2R can be easily expressed in terms of the properties of the junction cluster, 
\begin{equation}\label{a_eff}
\alpha_{eff}=P_{11}. 
\end{equation}
The effective exit rates from the left segments can also be found using similar arguments: the particle can leave the segment 1L or 2L  to the right if the vertical cluster is in the partially filled configuration, leading us to the following expression, 
\begin{equation}\label{b_eff}
\beta_{eff}=P_{10}=\frac{1-P_{11}}{2}.
\end{equation}
This result has  important consequences on the particle dynamics in the system. Since $\beta_{eff} \le 1/2$ the maximal-current phase cannot be realized in the left segments. Symmetry and conservation of the particle current also require that
\begin{equation}\label{current_sym}
J_{1L}=J_{2L}=J_{1R}=J_{2R},
\end{equation}
i.e., fluxes through all segments are equal.  Using the notation ($x$,$y$), where $x$ and $y$ denote the phase
of the left and right segments respectively,  we conclude that the only possible stationary phases for this system are (HD,HD), (LD,HD), (LD,LD) and (HD,LD) phases.

The (HD,HD) phase might exist if the following conditions are satisfied,
\begin{equation}\label{HDHD_cond}
\beta < 1/2, \quad \beta_{eff} < 1/2, \quad \beta_{eff} < \alpha, \quad \beta < \alpha_{eff}.
\end{equation}
Because the bulk densities in the left and right segments of this phase are the same, it implies that 
\begin{equation}
\beta_{eff}=\beta.
\end{equation}
Then from Eqs. (\ref{a_eff}) and (\ref{b_eff}) we derive that
\begin{equation}
\alpha_{eff}=1-2 \beta.
\end{equation}
Substituting this expression into inequalities (\ref{HDHD_cond}) we obtain, that (HD,HD) phase is realized when
\begin{equation}
\beta<\alpha, \quad \beta<\frac{1}{3}.
\end{equation}
Similar arguments for (LD,LD)  yield that this phase exists for 
\begin{equation}
\alpha <\beta, \quad \alpha<\frac{1}{3}.
\end{equation}
The conditions for existence of (HD,LD) phase can be written in the following form,
\begin{equation}\label{HDLD_cond}
\alpha_{eff}<1/2. \quad \beta_{eff}<1/2, \quad \alpha_{eff}<\beta, \quad \beta_{eff}<\alpha.
\end{equation}
In  this phase the stationarity of the particle current requires that $\alpha_{eff}=\beta_{eff}$, and utilizing Eqs.  (\ref{a_eff}) and (\ref{b_eff}) it produces
\begin{equation}
\alpha_{eff}=\beta_{eff}=\frac{1}{3}.
\end{equation}
Combining this result with Eq. (\ref{HDLD_cond}) we obtain that this phase exists only for
\begin{equation}
\alpha>1/3, \quad \beta > 1/3.
\end{equation}
It can also be shown that in this mean-field approximation that neglects the correlations between the segments the (LD,HD) phase correspond to the phase boundary 
\begin{equation}
\alpha = \beta < 1/3.
\end{equation}

In the general case of symmetric coupling between channels, when $w_1=w_2=w \le 1$, the calculations become slightly more involved because the vertical configuration with both empty sites now have  a nonzero probability, i.e., $P_{00} \ne 0$. The expressions for the effective entrance and exit rates are modified as
\begin{equation}\label{rates}
\beta_{eff}=1-P_{11}-P_{10}, \quad \alpha_{eff} = P_{11} +(1-w)P_{10}.
\end{equation}
As in the case of strong symmetric coupling, the particle fluxes through all segments are the same, and it is impossible to have a maximal-current phase in the system. If MC would exist in one of the segments, than it would have to be found in all segments. This would require that $\alpha_{eff} >1/2$ and $\beta_{eff} >1/2$, leading to
\begin{equation}\label{inequal}
\alpha_{eff}+\beta_{eff} >1.
\end{equation}
However, from Eq. (\ref{rates}) it follows that
\begin{equation}
\alpha_{eff}+\beta_{eff}=1-w P_{10}  <1,
\end{equation}
which contradicts the inequality (\ref{inequal}). Thus, our mean-field approach suggests that there are four possible stationary phases [(HD,HD), (LD,LD), (HD,LD) and (LD,HD)] in the two-channel ASEP with inhomogeneous symmetric coupling.

Let us define functions $\rho_{1}^{L}$ and  $\rho_{2}^{L}$ as probabilities to find the particle at the last sites of left segments 1L and 2L respectively. Similarly, $\rho_{i}^{R}$ ($i=1,2$) gives the density at the first sites of the right segments 2L and 2R. Because of the symmetry these functions are related,
\begin{equation}
\rho_{1}^{L}=\rho_{2}^{L}=\rho^{L}, \quad \rho_{1}^{R}=\rho_{2}^{R}=\rho^{R}.
\end{equation}
The dynamics at the vertical cluster is described at all times by the master equation,
\begin{equation}
\frac{d P_{11}}{d t}= \rho^{L}(P_{01}+P_{10})-2 P_{11}(1-\rho^{R}),
\end{equation}
which at steady state reduces to
\begin{equation}\label{master1}
\rho^{L}P_{10}= P_{11}(1-\rho^{R}).
\end{equation}
The particle current that leaves any of the left segments can be written as
\begin{equation}\label{JL}
J=\beta_{eff} \rho^{L},
\end{equation}
while the flux entering any of right segments is given by
\begin{equation}\label{JR}
J=\alpha_{eff}(1- \rho^{R}).
\end{equation}
Combining Eqs. (\ref{rates}), (\ref{master1}), (\ref{JL}) and   (\ref{JR}), we derive the expression  that relates the probabilities of filled and partially filled vertical junction  clusters,
\begin{equation}
(1-w) P_{10}^{2} +2 P_{11} P_{10} +P_{11}^{2} - P_{11}=0.
\end{equation}
Solving this equation and eliminating the unphysical solution leads us to
\begin{equation}
 P_{10}= \frac{\sqrt{P_{11}[1-w(1-P_{11})]}-P_{11}}{1-w}.
\end{equation}
Then the effective entrance and exit rates can be expressed using Eqs. (\ref{rates}) as
\begin{eqnarray}\label{rates1}
\alpha_{eff} &=& \sqrt{P_{11}[1-w(1-P_{11})]}, \nonumber \\
\beta_{eff} &=& \frac{(1-w)(1-P_{11})-\sqrt{P_{11}[1-w(1-P_{11})]}}{1-w}.
\end{eqnarray}

Analyzing the conditions for existence of four possible phase with the help of relations (\ref{rates1}), as was done above in the case of the strong coupling ($w=1$), it can be shown that (HD,LD) phase exists for 
\begin{equation}
\alpha>\frac{2-w}{4-w}, \quad \beta > \frac{2-w}{4-w}.
\end{equation}
Bulk density profiles and particle currents in this phase are
\begin{equation}
\rho_{bulk}^{L}=\frac{2}{4-w}, \quad \rho_{bulk}^{R}=\frac{2-w}{4-w}, \quad J_{1}=J_{2}=\frac{2(2-w)}{(4-w)^{2}}.
\end{equation}
(LD,LD) phase is found when
\begin{equation}
\alpha <\frac{2-w}{4-w}, \quad \alpha < \beta,
\end{equation}
and stationary-state properties are given by
\begin{equation}
\rho_{bulk}^{L}=\rho_{bulk}^{R}=\alpha, \quad J_{1}=J_{2}=\alpha(1-\alpha).
\end{equation}
Similarly, (HD,HD) phase exists for 
\begin{equation}
\beta <\frac{2-w}{4-w}, \quad \beta < \alpha.
\end{equation}
The bulk density and the particle flux in this phase are equal to
\begin{equation}
\rho_{bulk}^{L}=\rho_{bulk}^{R}=1-\beta, \quad J_{1}=J_{2}=\beta(1-\beta).
\end{equation}

The phase (LD,HD) is predicted to be found at the line $\alpha=\beta<\frac{2-w}{4-w}$. However, because our theory is a mean-field approach, it can be argued that in the real system it does not exist. At the phase transformation line a linear density profile, characteristic for first-order phase transitions, is expected \cite{derrida98}. The calculated phase diagram is also shown in figure 4. Note that for $w=0$ two channels are uncoupled and we recover the results for individual single-lattice ASEPs, as expected. In the limit of strong coupling ($w=1$) the predicted phase diagram is also agrees with the one discussed earlier. 

It is interesting to note that in the case of symmetric coupling  the phase diagram for two-channel ASEPs is very similar to a phase diagram for the single-chain ASEPs with local inhomogeneity \cite{kolomeisky98}. The connection between two systems can be easily understood from figure 1. In each lane the hoping rate to the right is the same at all sites except at the one that couples the channels. Then the motion along each lattice can be viewed as a transport along a single-lane ASEP with local defect  that was studied in Ref. \cite{kolomeisky98}.

\subsection{Asymmetric coupling}

The dynamics in the system of two-channel ASEPs with inhomogeneous coupling become very complex for  asymmetric vertical transition rates, i.e., when $w_{1} \ne w_{2}$., and the symmetry between two lattices no longer exists. Specifically, the densities at the last sites of the left segments and densities at the first sites of right segments are not equal: $\rho_{1}^{L} \ne \rho_{2}^{L}$, and $\rho_{1}^{R} \ne \rho_{2}^{R}$. To specify a state of the system we  use the notation $(x,y; w,z)$  with  $x$, $y$, $w$ and $z$ corresponding to the phase in the segments 1L, 1R, 2L and 2R respectively, and each of them is able to take the values LD, HD or MC. It is convenient to view the system with the asymmetric coupling slightly differently from the symmetric case, as shown in figure 3. Segments 2L and 1R have $L/2$ sites while 1L and 2R segments are  longer with $L/2+1$ sites. However, because we consider dynamics in the thermodynamic limit ($L \gg 1$), the difference in the sizes does not matter and all segments still can be viewed as homogeneous single-chain ASEP with known stationary properties.

\begin{figure}[h]
\centering \includegraphics[scale=0.6,clip=true]{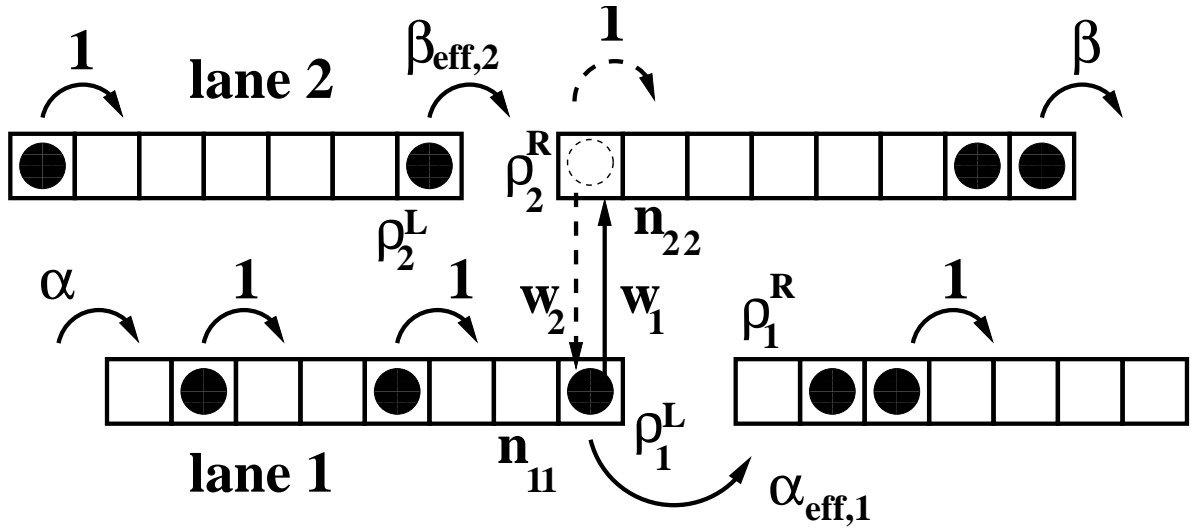} 
\caption{Effective scheme for two-channel ASEPs with inhomogeneous asymmetric coupling.}
\end{figure}

To simplify calculations, we consider the case of maximal asymmetry with $w_{1}=1$ and $w_{2}=0$. Our calculations can be easily generalized for general values of vertical transition rates. The particle can enter the segment 1R only when the vertical cluster is fully occupied, and it produces the following expression for the effective entrance rate,
\begin{equation}\label{alpha_eff}
\alpha_{eff,1}=P_{11}.
\end{equation}
However, as can be seen from figure 3, there are two ways to enter the segment 2R: from the left segment 2L and from the partially-filled  vertical configuration with the probability $P_{10}$. Then the effective entrance in this case is equal to 
\begin{equation}
\alpha_{eff,2}=\rho_{2}^{L}+ \frac{P_{10}}{1-\rho_{2}^{R}}, \quad \rho_{2}^{R}=P_{11}+P_{01}.
\end{equation}
Using similar arguments it can be shown that the effective exit rate from the segment 2L is given by the expression,
\begin{equation}\label{beta_eff}
\beta_{eff,1}=1- \frac{\rho_{1}^{R}P_{11}}{\rho_{1}^{L}}, \quad \rho_{1}^{L}=P_{11}+P_{10},
\end{equation}
while the effective rate out of 2L segment is 
\begin{equation}
\beta_{eff,2}=1-\rho_{2}^{R}.
\end{equation}

The densities at the last and at the first sites of segments play important role in our analysis. They can be easily expressed in terms of the entrance and exit rates from known exact solutions for single-chain ASEPs \cite{derrida98}. In the maximal-current  phase the densities are
\begin{eqnarray}\label{mc}
\rho_{1}^{L}=\frac{1}{4\beta_{eff,1}}, &  & \ \rho_{2}^{L}=\frac{1}{4\beta_{eff,2}},\nonumber \\
 \rho_{1}^{R}=1-\frac{1}{4\alpha_{eff,1}}, & & \ \rho_{2}^{R}=1-\frac{1}{4\alpha_{eff,2}}.
\end{eqnarray}
The corresponding expressions for segments in the low-density regimes are equal to
\begin{eqnarray}\label{ld}
\rho_{1}^{L}=\frac{\alpha(1-\alpha)}{\beta_{eff,1}}, & & \rho_{2}^{L}=\frac{\alpha(1-\alpha)}{\beta_{eff,2}},\nonumber \\
 \rho_{1}^{R}=\alpha_{eff,1}, & & \rho_{2}^{R}=\alpha_{eff,2};
\end{eqnarray}
while expressions in the high-density  regimes can be written as
\begin{eqnarray}\label{hd}
\rho_{1}^{L}=1-\beta_{eff,1} , & & \rho_{2}^{L}=1-\beta_{eff,2},\nonumber \\
 \rho_{1}^{R}=1-\frac{\beta(1-\beta)}{\alpha_{eff,1}}, & & \rho_{2}^{R}=1-\frac{\beta(1-\beta)}{\alpha_{eff,2}}.
\end{eqnarray}

Particle dynamics through vertical junction cluster can be described by a set of three independent master equations,
\begin{equation}
\frac{dP_{11}}{dt}= n_{11}P_{01}+\rho_2^LP_{10}-(1-\rho_1^R)P_{11}-(1-n_{22})P_{11};
\end{equation}
\begin{equation}
\frac{dP_{10}}{dt}=(1-\rho_1^R-n_{11})P_{11}-n_{11}P_{01}+n_{11}-(1+n_{11}+\rho_2^L)P_{10};
\end{equation}
\begin{equation}
\frac{dP_{01}}{dt}=(1-n_{22}-\rho_2^L)P_{11}+(1-\rho_2^L)P_{10}+\rho_2^L-(1-n_{22}+n_{11}+\rho_2^L)P_{01}; 
\end{equation}
where $n_{11}$ denotes the density at the site before the last one in the segment 1L and $n_{22}$ is the occupancy of the second site in the segment 2R: see figure 3. At steady-state these equations change  to the following expressions,
\begin{equation}\label{master_eq1}
n_{11}P_{01}+\rho_2^LP_{10} = (1-\rho_1^R)P_{11}+(1-n_{22})P_{11};
\end{equation}
\begin{equation}\label{master_eq2}
(1-\rho_1^R-n_{11})P_{11}+n_{11} =  n_{11}P_{01}+(1+n_{11}+\rho_2^L)P_{10}; 
\end{equation}
\begin{equation}\label{master_eq3}
(1-n_{22}-\rho_2^L)P_{11}+(1-\rho_2^L)P_{10}+\rho_2^L = (1-n_{22}+n_{11}+\rho_2^L)P_{01}.
\end{equation}

At large times the stationarity  in the system also  implies the following relations between the particle currents across different segments:
\begin{equation}\label{currents}
J^L_1+J^L_2=J^R_1+J^L_2, \quad J_1^R=J_1^L-P_{10}, \quad J_2^R=J_2^L+P_{10}.
\end{equation}
Since each segment can exist in one of three stationary phases (low-density, high-density or maximal-current), there are $3^{4}=81$  possible phases in the system. However, not all phases exist due to constraints of steady-state conditions. MC phase cannot exist in the segment 1R because from Eq. (\ref{currents}) we have  $J_{1}^{R} < J_{1}^{L}$ and 1/4 is the maximal possible flux through any segment. Similarly, it is not possible to have MC phase in the segment 2L since  $J_{2}^{L} < J_{2}^{R}$. These arguments already eliminate  $27+27-9=45$ phases. It is also not possible to have at stationary state the phases (HD,LD; $x$,$y$) for any $x$ and $y$. For these phase to exist we must have from Eqs. (\ref{alpha_eff}), (\ref{beta_eff}), (\ref{ld})  and (\ref{hd})   
\begin{equation}
\beta_{eff,1}=1-\frac{P_{11}^{2}}{P_{11}+P_{10}} \mbox{   and   } \beta_{eff,1}=1-P_{11}-P_{10}.
\end{equation}
Comparing these two equations leads to the expression $P_{10}^{2}+2 P_{11}P_{10}=0$, which cannot be satisfied at large times. In analogous way it can be shown that  phases ($x$,$y$; HD,LD) are not found at stationary state. Finally, using similar approach we have determined that there are only 10 phases that can be supported in the system at $t \rightarrow \infty$. 

The resulting phase diagram is presented in figure 5. Our theoretical calculations show that  (MC,LD;HD,MC) phase can be found for $\alpha > 1/2$ and $\beta >1/2$; (MC,LD;HD,HD) exists for  $\alpha > 1/2$ and $0.24<\beta<1/2 $; (MC,HD;HD,HD) phase  is realized for $\alpha>1/2$ and $0.22<\beta<0.24$; (LD,LD;HD,MC) phase  exists for  $0.24<\alpha<1/2$ and $\beta>1/2$; and (LD,LD;LD,MC) phase can be found for $0.19<\alpha<0.24$ and $\beta>1/2$. The curved boundaries between different phases are found by solving the appropriate master equations, and it yields that the curved border between  (LD,LD;LD,LD) and (LD,LD;LD,HD) phases is given by
\begin{eqnarray}
 \left[\beta+\frac{1}{2}(M-1)\right] \left[(1-\alpha)\alpha+\frac{1}{4}(2\alpha-1-2\alpha^2-2\beta+2\beta^2+M)^2 \right] &=& \nonumber \\
 -\frac{1}{4}(M-1)(3-2\beta^2+M) - \frac{\alpha(1-\alpha)(\alpha-\alpha^2-\beta+\beta^2)}{\beta-1},& &
\end{eqnarray}
where the  parameter $M$ is defined as
\begin{equation}
M=\sqrt{1-8\alpha+8\alpha^2+4\beta-4\beta^2}.
\end{equation}
The border between the (LD,LD;LD,HD) and (LD,LD;HD,HD) phases is calculated from the  equation
\begin{eqnarray}
\frac{1}{2} \left[(1-2\alpha+M)[(1-\alpha)\alpha+\frac{1}{4}(2\alpha-2\alpha^2-2\beta+2\beta^2+M-1)^2 \right]= \nonumber \\
(\alpha-1)(\alpha^2-\alpha+\beta-\beta^2)- \frac{1}{4}\left[(M-1)(1+2\alpha^2+4\beta-4\beta^2+M)\right].
\end{eqnarray}
The curved boundary between (LD,LD;HD,HD) and (LD,HD;HD,HD) phases is computed from
\begin{eqnarray}
(\alpha-2\alpha^3+\alpha^4+2\alpha\beta^2-2\alpha^2\beta^2+\beta^4)(1-2\beta+K) = &  & \nonumber \\
\beta(3+2\alpha-2\alpha^2-2\beta-K) - (\alpha-\alpha^2-\beta+\beta^2)(1+K), & &
\end{eqnarray} 
with 
\begin{equation}
K=\sqrt{1+4\alpha-4\alpha^2-8\beta+8\beta^2}.
\end{equation}
The border between (LD,HD;HD,HD) and (HD,HD;HD,HD) phases is described by
\begin{equation}
1+2\alpha^3+3\beta+\alpha^2(1-\beta+\beta^2)+\alpha(K-3)=\alpha^4+3\beta^2+K.
\end{equation}
Particle currents and density profiles in each phase can be found by substituting the appropriate solutions of master equations into  Eqs. (\ref{mc1}), (\ref{ld1}) and (\ref{hd1}).

\section{Monte-Carlo simulations and discussion}

The presented theoretical  description of two-channel ASEPs with inhomogeneous coupling treats the particle dynamics in horizontal segments and in the vertical junction cluster exactly. However, coupling between different parts of the system is viewed in a mean-field manner that neglects correlations. In order to test the validity of the theoretical analysis we performed extensive Monte Carlo computer simulations.   

\begin{figure}[h]
\centering \includegraphics[scale=0.4,clip=true]{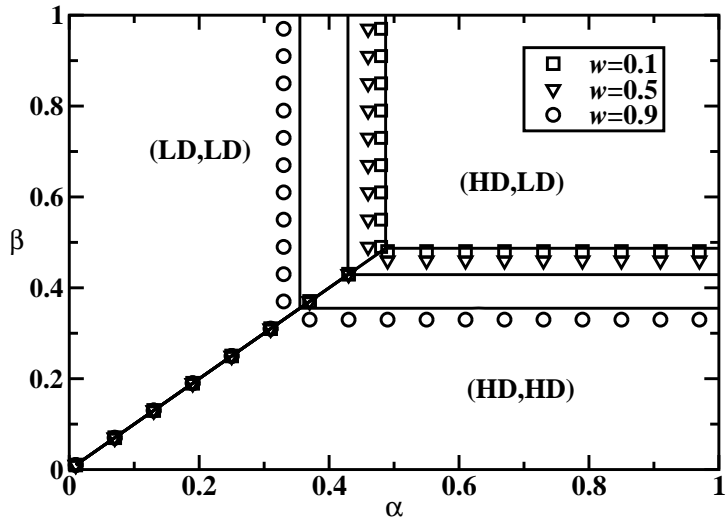} \caption{Phase diagram for two-channel ASEPs with different symmetric inhomogeneous couplings. Symbols correspond to Monte Carlo simulations, while lines are theoretical predictions. All phase transformations are first-order transitions with density jumps.}
\end{figure}

\begin{figure}[h]
\centering \includegraphics[scale=0.4,clip=true]{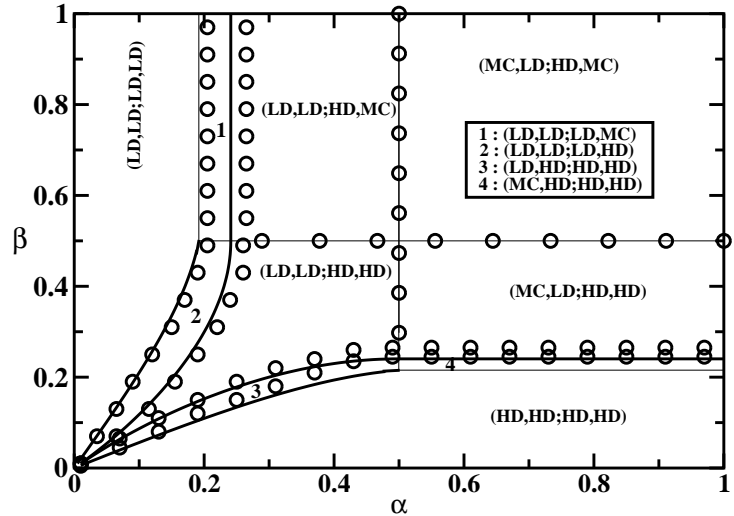} \caption{Phase diagram for two-channel ASEPs with asymmetric inhomogeneous coupling ($w_{1}=1$ and $w_{2}=0$). Symbols correspond to Monte Carlo simulations, while lines are theoretical predictions. Thick lines describe first-order phase transitions, thin lines represent continuous phase transformations.}
\end{figure}

To speed up computer simulations we utilized the so-called BKL algorithm \cite{BKL}. The total number of steps has been varied between $2\times 10^7$ and $10^9$. To ensure that our system had reached the stationary state we ignored first  $3-5\%$ of all steps in each simulation. Our theoretical results are valid in the thermodynamic limit of $L\rightarrow\infty$, indicating that finite-size effects are neglected. In our simulations we used mostly $L=1000$ sites lattices, but occasionally checked some computations for lattices of sizes $L=100$ or $L=10000$.  Computed dynamic properties have been found to be  unaffected by lattice size variation, confirming absence of finite-size effects. Phase diagram boundaries between different phases have been determined by analyzing density profile and particle currents. In the case where we have transitions between LD and HD phases it is possible to visually observe  the phase boundary: it happens when the density profile becomes linear. This method allowed us to determine the phase boundaries  within 0.01 units of the numerical values assigned to $\alpha$ and $\beta$. In the case of transitions between an MC phase and LD/HD phases, phase borders have been determined  by observing the saturation of the particle current for specific $\alpha$ and $\beta$ values. This approach, although not as exact as the previous one, still allows us to compute the phase transition line to within 0.05 units of the $\alpha$ and $\beta$ values.

\begin{figure}[h]
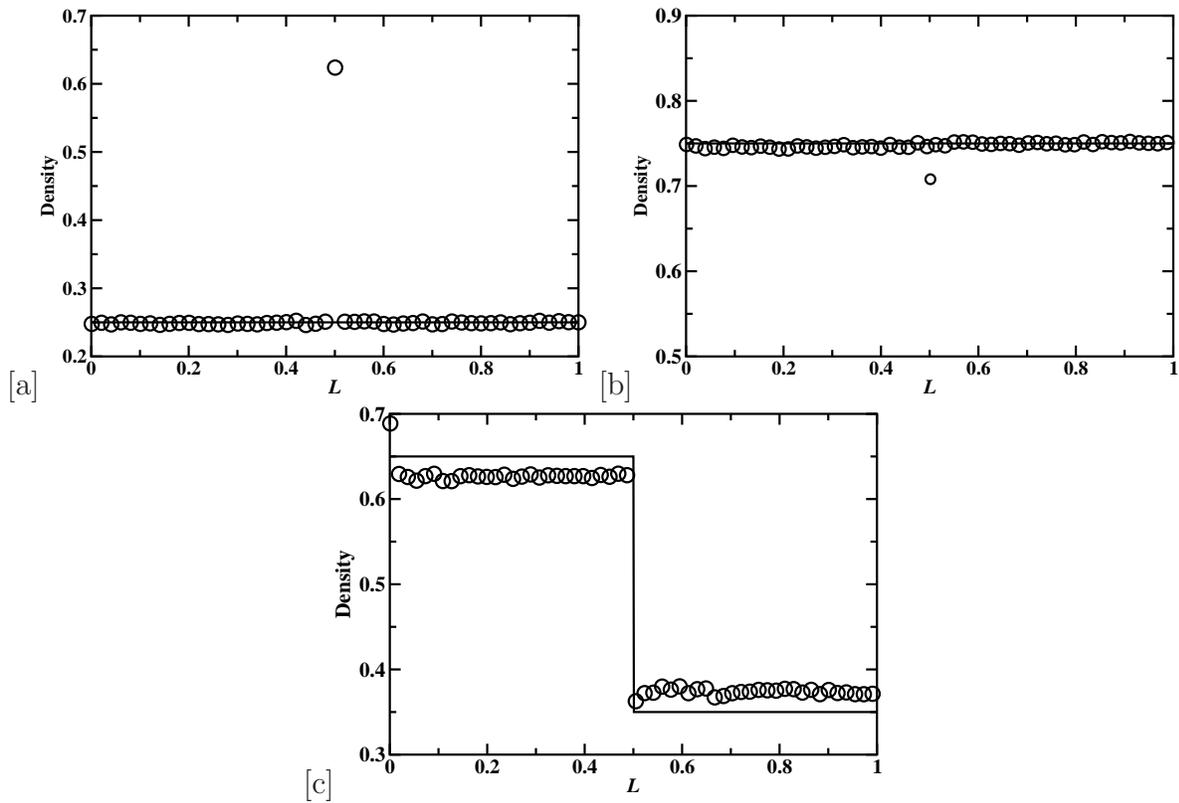

\centering [a]\includegraphics[scale=0.3,clip=true]{Fig6a.eps}
           [b]\includegraphics[scale=0.3,clip=true]{Fig6b.eps}
           [c]\includegraphics[scale=0.3,clip=true]{Fig6c.eps} 
\caption{Density profiles for two-channel ASEPs with symmetric inhomogeneous couplings: a) (LD,LD) phase with $w=1$, $\alpha=0.25$ and $\beta=0.75$; b) (HD,HD) phase with $w=0.5$, $\alpha=0.75$ and $\beta=0.25$; c) (HD,LD) phase with $w=0.9$ and $\alpha=\beta=0.75$. Symbols correspond to Monte Carlo simulations, while lines are theoretical predictions for bulk densities.}
\end{figure}

Phase diagram for two-channel ASEPs with inhomogeneous symmetric coupling, obtained from Monte Carlo simulations is presented in figure 4, and it can be concluded that the approximate theoretical approach agrees well with computer simulations. The agreement is excellent for weak coupling between the channels ($w \ll 1$), while increasing the probability of vertical transitions  ($w \simeq 1$) leads to stronger correlations in the system, and theoretical predictions start to deviate from Monte Carlo simulations. More information can be obtained from density profiles plotted in figure 6. Our theory gives excellent agreement for bulk densities in (HD,HD) and (LD,LD) phases for any coupling between the channels: see figures 6a and 6b. This is because the dynamics of the system in these cases is governed by entrance or exit processes. However, in (HD,LD) phase the theoretical predictions only qualitatively agree with computer simulations, as can be seen in figure 6c. This observation can be explained by the fact that in this phase the dynamics is governed by processes near the vertical junction cluster, and our theory neglects correlations between different segments of the system.

\begin{figure}[h]
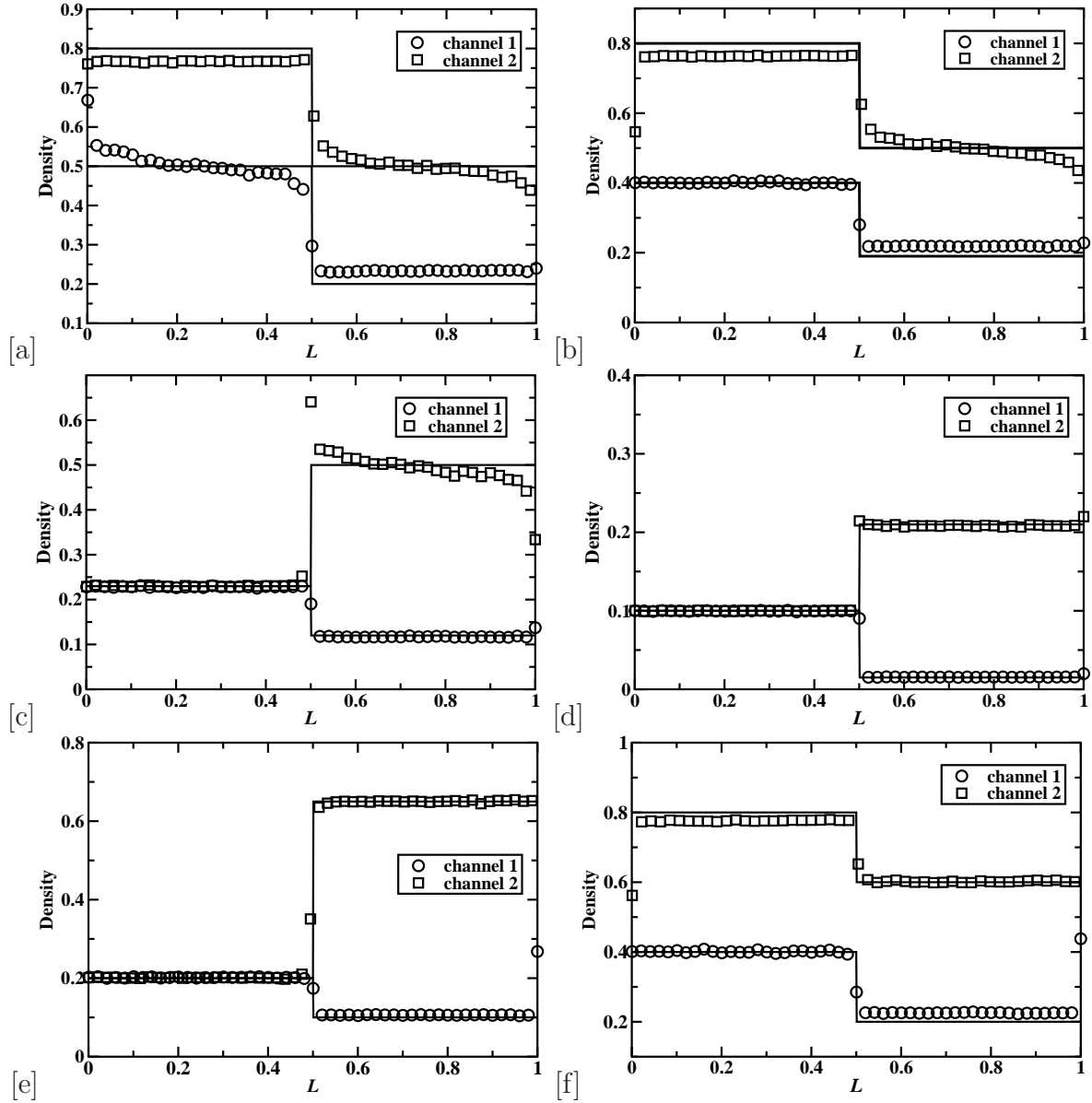

\centering [a]\includegraphics[scale=0.3,clip=true]{Fig7a.eps}
           [b]\includegraphics[scale=0.3,clip=true]{Fig7b.eps}
           [c]\includegraphics[scale=0.3,clip=true]{Fig7c.eps} 
           [d]\includegraphics[scale=0.3,clip=true]{Fig7d.eps} 
           [e]\includegraphics[scale=0.3,clip=true]{Fig7e.eps} 
           [f]\includegraphics[scale=0.3,clip=true]{Fig7f.eps}
\caption{Density profiles for two-channel ASEPs with asymmetric inhomogeneous couplings ($w_{1}=1$ and $w_{2}=0$): a) (MC,LD; HD,MC) phase with $\alpha=\beta=0.75$; b)  (LD,LD; HD,MC) phase with $\alpha=0.4$ and $\beta=0.75$; c) (LD,LD; LD,MC) phase with $\alpha=0.23$ and $\beta=0.75$; d) (LD,LD; LD,LD) phase with $\alpha=0.1$ and $\beta=0.75$; e) (LD,LD; LD,HD) phase with $\alpha=0.2$ and $\beta=0.35$;  f) (LD,LD; HD,HD) phase with $\alpha=\beta=0.4$. Symbols correspond to Monte Carlo simulations, while lines are theoretical predictions for bulk densities.}
\end{figure}

Stationary-state phase behavior in two-channel ASEPs with inhomogeneous asymmetric coupling is much more complex, as shown in figure 5. Monte Carlo computer simulations suggest that there are 10 stationary phases, in agreement with theoretical predictions. Theoretical calculations for phase boundaries mostly agree  with computer simulations, although there are deviations for small $\alpha$ and/or small $\beta$. Density profiles of six phases for asymmetric coupling are presented in figure 7.   The agreement between theoretical calculations and computer simulations is excellent in  the phases where dynamic properties depend on the entrance and exit rates: see figures 7c, 7d and 7e. However, deviations between the theory and Monte Carlo simulations are observed in the phases where dynamics is specified by the processes near the vertical junction cluster. Although our theoretical calculations and Monte Carlo computer simulations have been performed for the case of strong asymmetry ($w_{1}=1$ and $w_{2}=0$), they can be easily extended for other asymmetric transition rates, and we expect a similar phase diagram and stationary densities and particle currents.

Our theoretical calculations, supported by Monte Carlo computer simulations, indicate that for symmetric coupling between the channels the phase diagram is quite simple with only three phases. However, breaking this symmetry leads to a very complex dynamics with 10  stationary-state phases. This behavior can be understood by analyzing particle fluxes through segments. In the symmetric coupling the currents in all segments are the same. In addition, it was shown that the maximal-current phase cannot exist under these conditions. These constraints lead to a simple phase diagram as outlined in the figure 4. In contrast, for the asymmetric coupling between the channels the currents through all segments are not necessary equal to each other, and MC phase can be found under some conditions in some segments. This leads to existence of many steady-state phases in the system. 

The phase diagram for the asymmetric inhomogeneous coupling is almost symmetric with respect to $\alpha \leftrightarrow \beta$ transformation. This is because the transport of particles from the left can be also viewed as a motion of holes in the opposite direction. However, as can be clearly seen from figure 5 there is a slight asymmetry in the phase diagram, and it also shows up in the dynamic properties. The deviation from the particle-hole symmetry can be explained by analyzing the dynamics at the vertical junction for the case of $w_{1}=1$ and $w_{2}=0$, as illustrated in figure 8. The particle at the partially filled junction cluster can only move in the vertical direction. At the same time, the hole in the partially filled cluster can move vertically {\it and} horizontally, if the particle from the 1L segments enters the cluster. Thus, the corresponding  dynamic rules for the particles and for the holes in the vertical cluster are not the same, and it produces the slight asymmetry in large-time properties of two-channel ASEPs with inhomogeneous asymmetric coupling.

\begin{figure}[h]
\centering \includegraphics[scale=0.5,clip=true]{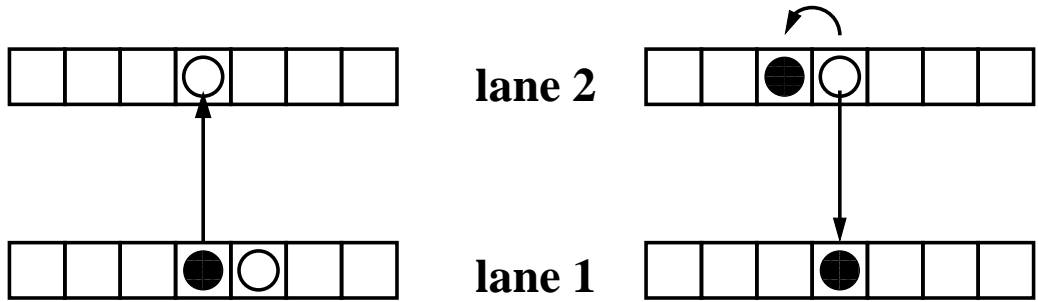} \caption{Dynamic transitions for the particle (on the left) and for the hole (on the right) at the vertical cluster. Allowed transitions are shown by arrows.}
\end{figure}
\vspace{3 cm}

\section{Summary and conclusions}

Two-channel asymmetric simple  exclusion processes with inhomogeneous coupling between the lattices have been investigated. To better describe the large-time behavior of the system a mean-field approach, that treats particle dynamics in four lattice segments and in the vertical cluster exactly while neglecting the correlations between different parts of the system, is developed. For symmetric transition rates between the channels the phase diagram is rather simple and it consists of three phases. This is a consequence of constraints on the particle fluxes due to the symmetry. The situation is very different for asymmetric coupling between the channels. Here our analysis indicates that there are ten stationary phases. It is argued that this complex behavior is realized because of weaker constraints on the particle currents. Our theoretical calculations are supported by extensive Monte Carlo computer simulations. Slight deviations between theoretical results and computer simulations are only found in the phases where correlations around the junction site play important role.  Our theoretical calculations and computer simulations also point out to asymmetry in the properties of two-channel ASEPs with asymmetric coupling with the respect to the particle-hole symmetry. It is argued that this asymmetry appears due to different dynamic rules for the particles and for the holes at the vertical junction cluster.

There are several extensions of this work can be proposed. Two-channel ASEPs with homogeneous asymmetric coupling show a phase diagram with only seven phases and very different stationary properties \cite{pronina07}. It will be interesting to increase the region of inhomogeneous coupling from one site to many to see how dynamics of the system changes.  Another interesting question that will help to understand  mechanisms of asymmetric exclusion processes  is how far coupling clusters should be put in order to feel each other. In addition, the analysis can be extended to transport of extended particles \cite{chou04}.

\section{Acknowledgments}The authors acknowledge the support from the Welch Foundation (under grant no. C-1559) and from the US  National Science Foundation through the grant CHE-0237105. K.Tsekouras would like to thank E.Pronina and S.Kotsev for their valuable help with computational issues.


\section*{References}

\begin{thebibliography}{99}

\bibitem{derrida98} Derrida B 1998 {\it Phys. Rep} {\bf 301} 65

\bibitem{schutz} Sch\"{u}tz G M 2000 Integrable stochastic many-body systems {\it Phase Transitions and Critical Phenomena} vol 19, ed
Domb  and Lebowitz   (London: Academic)

\bibitem{schutz03} Sch\"{u}tz G M 2003  {\it J. Phys. A: Math. Gen.}, {\bf 36} R339

\bibitem{macdonald68} MacDonald J T, Gibbs J H and Pipkin A C 1968 {\it Biopolymers} {\bf 6} 1

\bibitem{shaw03} Shaw L B, Zia R K P and Lee K H 2003 {\it Phys. Rev. E}, {\bf 68} 021910

\bibitem{shaw04} Shaw L B,  Kolomeisky A B and  Lee K H 2004 {\it J. Phys. A.: Math. Gen.}, {\bf 37} 2105

\bibitem{chou04} Chou T and Lakatos G 2004 {\it Phys. Rev. Lett.},  {\bf 93}  198101

\bibitem{chou99} Chou T and Lohse D 1999 {\it Phys. Rev. Lett.},  {\bf 82} 3552

\bibitem{klumpp05} Klumpp S, Nieuwenhuizen T M and Lipowsky R 2005 {\it Physica E}, {\bf 29} 380

\bibitem{parmeggiani03} Parmeggiani A, Franosch T and Frey E 2003 {\it Phys. Rev. Lett.},  {\bf 90} 086601

\bibitem{nishinari05} Nishinari K, Okada Y, Schadschneider A and Chowdhury D 2005 {\it Phys. Rev. Lett.},  {\bf 95} 118101

\bibitem{evans07} Evans M R and Sugden K E P 2007 {\it Physica A}, {\bf 384} 53

\bibitem{helbing01} Helbing D 2001 {\it Rev. Mod. Phys.}, {\bf 73} 1067

\bibitem{chowdhury00} Chowdhury D, Santen L and Schadschneider A 2000 {\it Phys. Rep.}, {\bf 329} 199

\bibitem{popkov01} Popkov V and Peschel I 2001 {\it Phys. Rev. E}, {\bf 64} 026126

\bibitem{pronina04} Pronina E and Kolomeisky A B 2004 {\it J. Phys. A: Math. Gen.}, {\bf 37} 9907

\bibitem{pronina05} Pronina E and Kolomeisky A B 2005 {\it J. Stat. Mech.}, P07010

\bibitem{mitsudo05} Mitsudo T and Hayakawa H 2005 {\it J. Phys. A: Math. Gen.}, {\bf 38} 3087

\bibitem{harris05} Harris R J and  Stinchcombe R B 2005 {\it Physica A}, {\bf 354} 582

\bibitem{pronina06} Pronina E and Kolomeisky A B 2006 {\it Physica A}, {\bf 372} 12

\bibitem{reichenbach06} Reichenbach T, Franosch T  and Frey E 2006 {\it Phys. Rev. Lett.},  {\bf 97}  050603

\bibitem{pronina07} Pronina E and Kolomeisky A B 2007 {\it J. Phys. A: Math. Gen.}, {\bf 40} 2275

\bibitem{solomovici97}  Solomovici J, Lesnik T and Reiss C 1997 {\it J. Theor. Biol.}, {\bf 185} 511

\bibitem{kolomeisky98} Kolomeisky A B 1998 {\it J. Phys. A: Math. Gen.}, {\bf 31} 1153

\bibitem{mirin03}  Mirin N and  Kolomeisky A B  2003 {\it J. Stat. Phys.}, {\bf 110} 811

\bibitem{ha03} Ha M, Timonen J and den Nijs M 2003 {\it Phys. Rev. E}, {\bf 68} 056122

\bibitem{howard} Howard J 2001 {\it Mechanics of Motor Proteins and the Cytoskeleton} (Sunderland, Massachusetts: Sinauer Associates)

\bibitem{BKL} Bortz A B, Kalos M H, Lebowitz J L 1975 {\it J. Comp. Phys.}, {\bf 17} 10.





\end{thebibliography}
\end{document}